\documentclass[12p]{article}

\begin{document}

\input amssym.tex

\title{On  anti-de Sitter oscillators and de Sitter anti-oscillators} 

\author{Ion I.  Cot\u aescu \thanks{E-mail:~~~icotaescu@yahoo.com}\\
{\small \it West University of Timi\c soara,}\\
       {\small \it V.  P\^ arvan Ave.  4, RO-300223 Timi\c soara, Romania}}

\maketitle

\begin{abstract}
We revisit the principal arguments for interpreting the free quantum fields on anti-de Sitter or de Sitter spacetimes of any dimensions as oscillators or respectively anti-oscillators. In addition, we point out that there exists a chart on the de Sitter background  where the free Dirac field becomes a genuine anti-oscillator in the non-relativistic limit in the sense of special relativity.

Pacs:
04.62.+v
\end{abstract}


\newpage
The harmonic oscillators of frequency $\omega$ are among the most simple and popular classical or quantum systems. Changing $\omega\to i\omega$ one obtains other systems that are called often anti-oscillators. For example,  in the non-relativistic quantum mechanics on a $d$-dimensional Euclidean space of coordinates $\vec{x}=(x^1,x^2,...,x^d)$ the well-known Schr\" odinger equations of the harmonic isotropic oscillator and its associated anti-oscillator  of the same mass $m$ read
\begin{equation}\label{Sc}
\left[-\frac{1}{2m}\Delta +\frac{m\omega^2}{2} {\vec{x}}^2\right] \phi=i\partial_{t} \phi
\begin{array}{c}
\\
\longleftrightarrow\\{\omega\to i\omega}
\end{array}
\left[-\frac{1}{2m}\Delta -\frac{m\omega^2}{2} {\vec{x}}^2\right] \phi=i\partial_{t} \phi\,.
\end{equation}

In special relativity there are oscillating systems formed by scalar \cite{O1,O2,O3,O4} or spinor \cite{S1,S2} quantum fields  but whose field equations are not covariant. Thus it seems that the natural relativistic oscillators  are those of general relativity constituted by classical or quantum fields defined  on  anti-de Sitter (AdS) spacetimes \cite{A1,A2,A3,A4,A5}. Other systems simulating oscillatory motions in general relativity were studied in Refs. \cite{C1,CC1}. Moreover,  the terms of oscillator and anti-oscillator  are used today even in cosmology related to actual models with positive or negative cosmological constants \cite{Lem}. This suggests that typical system of general relativity are the AdS oscillators and the corresponding de Sitter (dS) anti-oscillators. 

Under such circumstances we believe that it is worth reviewing the properties of the quantum fields defined on these spacetimes that emphasise their characters of oscillator or anti-oscillator. For this reason, we would like to revisit here the principal features of these systems  considering the generalization to arbitrary dimensions \cite{CScalar,CAdS,CAdSS} of  some of our previous results \cite{COT,COT1,COT2}. In addition, we show that there exists a dS chart where the quantum fields behaves as genuine anti-oscillators in the non-relativistic limit.  We focus on the free Dirac quantum field on the mentioned manifolds since  this is the only quantum  field having the same rest energy as in special relativity \cite{COT3}. Why this property is important will be pointed out by comparing the Dirac field with the scalar one.

We discuss simultaneously both the manifolds under consideration here denoting the dS and AdS $d$-dimensional spacetimes by $(M^{\epsilon}_{1+d},g^{\epsilon})$ understanding that  $\epsilon=1$ corresponds to the dS case and $\epsilon=-1$ to the AdS one. These  spacetimes are hyperboloids embedded in the  $(d+2)$-dimensional flat spacetimes, $(M^{\epsilon}_{1+d+1},\eta^{\epsilon})$, of coordinates  $z^{A},\, A,B,...=0,1,2,...,d+1$ and metric
\begin{equation}
\eta^{\epsilon}={\rm diag}(1,\,\underbrace{-1,-1,...,-1}_{d},-\epsilon)\,.
\end{equation} 
The equations $\eta_{AB}^{\epsilon}z^{A}(x)z^{B}(x)=-\epsilon\,R^{2}$ of the hyperboloids of
radius $R=1/\omega$,  are solved by the functions $z^A(x)$ that introduce the coordinates $x^{\mu}$  ($\alpha,...\mu,...=0,1,2,...,d$) of the charts $\{x\}$  on these manifolds.

On  $(M^{\epsilon}_{1+d},g^{\epsilon})$ one defines the  {\em static} charts of  Cartesian coordinates,  $\{t,\vec{x}\}$ (with $t=x^0$), given by the parametrization    
\begin{eqnarray}
z^{0}\!\!&=&\omega^{-1}\chi_{\epsilon}(r)\left\{\begin{array}{lll}
\sinh \omega t&{\rm if}&\epsilon=1\\
\sin \omega t&{\rm if}&\epsilon=-1
\end{array}\right.\nonumber\\
z^{i}&=& x^{i}\,,\quad i=1,2,...,d\label{dScart}\\ 
z^{d}\!\!&=&\omega^{-1}\chi_{\epsilon}(r)\left\{\begin{array}{lll}
\cosh \omega t&{\rm if}&\epsilon=1\\
\cos \omega t&{\rm if}&\epsilon=-1
\end{array}\right.\nonumber
\end{eqnarray} 
where we denote by $r=|\vec{x}|$ the Euclidean norm of $\vec{x}$ and  
\begin{equation}\label{chi}
\chi_{\epsilon}(r)=\sqrt{1-\epsilon\,\omega^{2}r^2}\,.
\end{equation}
In the associated spherical coordinates, $\vec{x}\to (r,\theta_1,\theta_2,...,\theta_{d-1})$, defined as \cite{T} 
\begin{eqnarray}
x^1&=&r\cos\theta_1\sin\theta_2...\sin\theta_{d-1}\,,\nonumber\\
x^2&=&r\sin\theta_1\sin\theta_2...\sin\theta_{d-1}\,,\label{sferic}\\
&\vdots&\nonumber\\
x^d&=&r\cos\theta_{d-1}\,,\nonumber
\end{eqnarray}
the line elements  of these charts take the form
\begin{eqnarray}\label{dSmet}
ds^{2}&=&\eta_{AB}^{\epsilon}dz^{A}dz^{B}\label{(adsm)}\\
&=&\chi_{\epsilon}(r)^{2} dt^{2}-
\frac{dr^{2}}{\chi_{\epsilon}(r)^{2}} -
 r^{2}d\Omega_{d-1}^2\nonumber
\end{eqnarray}
where  $d\Omega_{d-1}^2={d\theta_1}^2+\sin^2\theta_1 {d\theta_2}^2+
\sin^2\theta_1 \sin^2\theta_2{d\theta_3}^2 +\cdots$ is the usual line element on the sphere $S^{d-1}$ \cite{T}. The radial domains of these charts are $D_{r}=[0,1/\omega)$ for the dS spacetimes and  $D_{r}=[0,\,\infty)$ for the AdS ones. 

The Dirac theory in curved spacetimes can be built starting with spaces of $2^k$-dimensional spinors, $\psi$,  where one can define $2k+1$ point-independent $\gamma$-matrices such that  we must take $k=\frac{d}{2}$  when $d$ is even or $k= \frac{d+1}{2}$ if $d$ is odd \cite{SG,CV}. The Dirac equation for a free  fermion of mass  $m$,   minimally coupled to the gravity of the background,
\begin{equation}\label{Dirac}
(i\gamma^{\mu}(x)\nabla_{\mu}-m)\psi(x)=0\,,\quad
\gamma^{\mu}(x)=e^{\mu}_{\hat\alpha}(x)\gamma^{\hat\alpha}\,,
\end{equation}
is defined in local frames, $\{t,\vec{x};e\}$, given by the gauge fields (or "vielbeins") $e_{\hat\alpha}$ and $\hat e^{\hat\alpha}$,  labelled by the local indices $\hat\alpha,...,\hat\mu,..$, having the same range as the natural ones.  The covariant derivatives having the action $\nabla_{\mu}\psi=(\partial_{\mu}+\Gamma_{\mu}^{spin})\psi$
depend on the spin connections denoted here as in Ref. \cite{CV}.

The free Dirac equation on the central charts defined above  can be brought in a simpler form by fixing the Cartesian  gauge defined  by the following 1-forms  in Cartesian coordinates \cite{CAdS,CAdSS}
\begin{eqnarray}
\tilde\omega^0&=&\hat e^0_{\mu}dx^{\mu}=\chi_{\epsilon}(r)\,dt\,,\\
\tilde\omega^i&=&\hat e^i_{\mu}dx^{\mu}=dx^i+\frac{1}{r^2}\left[\frac{1}{\chi_{\epsilon}(r)}-1\right]x^i \vec{x}\cdot d\vec{x}\,.
\end{eqnarray}
In this gauge the spherical variables of the free Dirac equation can be separated as in the spherical problems of special relativity, in terms of the angular spinors  $\phi_{\kappa,(j)}(\theta)$ defined in Ref.\cite{XYGu}. These depend only on the angular variables $\theta=(\theta_1,\theta_2,...,\theta_{d-1})$  and are determined by a set of weights $(j)$ of an irreducible representation of the group ${\rm Spin}(d)$ (i. e. the universal covering group of the group $SO(d)$) and the eigenvalue $\kappa=\pm |\kappa|$ of an operator which concentrates all the angular operators of the Dirac one \cite{CAdS}. 

The separation procedure is complicated being different for even or odd values of $d$ \cite{XYGu}. For odd $d$ the  particular solution of given energy, $E$, and positive frequency corresponding to the irreducible representation $(j)$ may have the form
\begin{equation}\label{(psol)}
\psi^{\epsilon}_{E,\kappa,(j)}(t,r,\theta)=r^{-\frac{d-1}{2}}\chi_{\epsilon}(r)^{-\frac{1}{2}}e^{-iEt}\left(
\begin{array}{c}
f^{\epsilon(+)}_{E,\kappa}(r)\phi_{\kappa,(j)}(\theta)\\
i f^{\epsilon(-)}_{E,\kappa}(r)\phi_{-\kappa,(j)}(\theta)
\end{array}\right)\,,
\end{equation}
where $\kappa$ can take  positive or negative  values. However, if $d$ is even then 
there are pairs of associated irreducible representations, $(j_1)$ and $(j_2)$, 
giving the same eigenvalue of the first Casimir operator, among them one takes 
the first one for positive values of $\kappa$ and  the second one for the 
negative values, $\kappa=-|\kappa|$. Consequently, the particular solutions 
read
\begin{eqnarray}
\psi_{E,|\kappa|,(j_1)}(t,r,\theta)&=&r^{-\frac{d-1}{2}}\chi_{\epsilon}(r)^{-\frac{1}{2}}e^{-iEt}\nonumber\\
&\times& \left[f^{\epsilon(+)}_{E,|\kappa|}(r)\phi_{|\kappa|,(j_1)}(\theta)
+if^{\epsilon(-)}_{E,-|\kappa|}(r)\phi_{-|\kappa|,(j_1)}(\theta)\right]\,,
\label{sol1}\\
\psi_{E,-|\kappa|,(j_2)}(t,r,\theta)&=&r^{-\frac{d-1}{2}}\chi_{\epsilon}(r)^{-\frac{1}{2}}e^{-iEt}\nonumber\\
&\times& \left[f^{\epsilon(+)}_{E,-|\kappa|}(r)\phi_{-|\kappa|,(j_2)}(\theta)
+if^{\epsilon(-)}_{E,|\kappa|}(r)\phi_{|\kappa|,(j_2)}(\theta)\right]\,.
\label{sol2}
\end{eqnarray}
It is remarkable that both these types of solutions involve  the same type  of radial wave functions  that depend on $E$ and   
\begin{equation}\label{kappa}
\kappa=\pm\left(\textstyle\frac{d-1}{2}+l\right)\,,\quad l=0,1,2,... \,,  
\end{equation}
where $l$ is an auxiliary orbital quantum number related to the representations $(j)$ or $(j_1)$ and $(j_2)$ \cite{XYGu}. These radial functions satisfy the system of radial equations
\begin{equation}\label{(e1)}
\left[\pm\chi_{\epsilon}(r)\frac{d}{dr}+\frac{\kappa}{r}\right]f^{\epsilon (\pm)}_{E,\kappa}(r)=\left[\frac{E}{\chi_{\epsilon}(r)}\pm m
\right]f^{\epsilon (\mp)}_{E,\kappa}(r)\,,
\end{equation}
that depend only on our function (\ref{chi}).

Thus we see that the free Dirac fields on dS or AdS spacetimes satisfy similar radial equations that can be transformed between themselves by replacing $\omega\to i\omega$. On the other hand, the energy eigenspinors on the AdS spacetime are square integrable corresponding to the discrete energy spectrum \cite{CAdS},
\begin{equation}\label{En}
E_n=m+\omega\left(n+\frac{d}{2}\right)\,, \quad n=0,1,2,...
\end{equation}   
which, in our opinion, defines the ideal isotropic harmonic oscillator of general relativity. We remind the reader that the free massive scalar field of mass $m$ minimally coupled to the AdS gravity has also equidistant energy levels  \cite{CScalar},
\begin{equation}\label{Ensc}
E_n^{sc} =M_{AdS}+\omega\left(n+\frac{d}{2}\right)\,, \quad M_{AdS}=\sqrt{m^2+\frac{d^2}{4}\omega^2}\,,
\end{equation}
with the same energy quanta but having  the ground state energy given by the  effective AdS mass $M_{AdS}$ instead of $m$.  This difference is due to the fact that in AdS  spacetimes the free Dirac equation is no longer the square root of the Klein-Gordon equation with the same mass.   Anyway, both these fields behave as oscillators such that it is natural to say that the corresponding fields on the dS spacetime behave as  anti-oscillators. 

Apart from this indirect argument, we can show that at least the free Dirac field on the dS spacetimes becomes a genuine anti-oscillator in the non-relativistic limit. This can be done by changing the central dS chart $\{t,\vec{x}\}$  into the de Sitter-Painlev\' e one $\{t',\vec{x}\}$ \cite{P1,P2} having the time
\begin{equation}\label{tt}
t'=t+\frac{1}{2\omega}\ln (1-\omega^2 r^2)
\end{equation}
and the line element 
\begin{equation}
ds^2=(1-\omega^2 {\vec{x}}^2) dt^{\prime\, 2}+2\omega \vec{x}\cdot d\vec{x} dt'- {d\vec{x}}^2\,.
\end{equation}
Here we  choose the Cartesian gauge defined by the 1-forms
\begin{eqnarray}
\tilde\omega^0&=&dt'\,,\\
\tilde\omega^i&=&dx^i-\omega x^i dt'\,.
\end{eqnarray}
In this gauge the Dirac equation  takes the  Hamiltonian form, 
\begin{equation}\label{D1}
\left[-i\gamma^0{\gamma}^i\partial_i+\gamma^0 m +{\cal H}_{int}\right]\psi=i\partial_{t'}\psi\,,
\end{equation}
having the traditional Minkowskian kinetic part and the  specific interaction term with the gravitational field of the dS manifold that reads,
\begin{equation}\label{Hint}
{\cal H}_{int}=-i\omega\left(x^i\partial_i+\frac{d}{2}\right)\,.
\end{equation}
Hereby we can derive the non relativistic limit (in the sense of special relativity)  obtaining the  Schr\" odinger equation $\left[-\frac{1}{2m}\Delta +{\cal H}_{int}\right]\phi=i\partial_{t'} \phi$,
of a non-relativistic massive particle (of mass $m$) freely moving  on the dS spacetime.  Furthermore,  we observe that the substitution $\phi\to e^{-\frac{i}{2}m\omega{\vec{x}}^2}\phi$
leads to Eq. (\ref{Sc}b) of a genuine non-relativistic  isotropic anti-oscillator. 

In the same chart $\{t',\vec{x}\}$, the example of the scalar field minimally coupled to the dS gravity  is less convincing since the non-relativistic limit of the Klein-Gordon equation, 
\begin{equation}
\left[(i\partial_{t'}-{\cal H}_{int})^2+\Delta-m^2+\textstyle\frac{d^2}{4}\omega^2\right]\phi=0\,,
\end{equation}
is a Schr\" odinger equation of the form  (\ref{Sc}b) but depending on  the effective dS mass,  
\begin{equation}
M_{dS}=\sqrt{m^2-\frac{d^2}{4}\omega^2}\,,
\end{equation}
that does not make sense for $m\le \frac{d}{2}\omega$.

In other respects, we must specify that in the AdS geometries there are no charts similar to the de Sitter-Painlev\' e  one since this might have a complex time resulted from the transformation (\ref{tt}) with $i\omega$ instead of $\omega$.  Therefore the equation resulted in this case,  
\begin{equation}\label{D2}
\left[-i\gamma^0{\gamma}^i\partial_i+\gamma^0 m +\omega\left({x}^i\partial_i+\frac{d}{2}\right)\right]\psi=i\partial_{t'}\psi
\end{equation}  
remains a mere mathematical problem of academic interest  but without a reasonable physical meaning (at least in this conjuncture). 

The conclusion is that the crucial argument for defining the Dirac AdS oscillator is its discrete energy spectrum (\ref{En}). The Dirac field on the dS spacetime can be considered as the corresponding anti-oscillator since  these two problems transform into each other by changing $\omega\to i\omega$. Moreover, the dS Dirac anti-oscillator lays out its nature in the non relativistic limit.  The scalar field has similar properties but with the difference that in this case the effective masses $M_{AdS}$ and $M_{dS}$ take over the role of the proper mass $m$.  

Finally, we note that the Dirac oscillators of general relativity we discussed here are completely different from those of special relativity obeying Dirac equations with supplemental terms linear in space coordinates \cite{S1,S2}. The argument is that on AdS or dS backgrounds the fermions satisfy free Dirac equations that in the flat limit ($\omega\to 0$)  give the free Dirac quantum modes of special relativity, losing thus their oscillator or anti-oscillator properties.

\end{document}